\documentclass[aps,prx,reprint,amsmath,amssymb,longbibliography,floatfix]{revtex4-2}
\ifdefined\pdfminorversion\pdfminorversion=7\fi
\usepackage[T1]{fontenc}
\usepackage{graphicx}
\usepackage{bm}
\usepackage{hyperref}
\hypersetup{hidelinks,pdftitle={Colloidal shadows reveal hidden solute transport},pdfauthor={Haoyu Liu, Zehao Chen, Amir A. Pahlavan}}

\begin{document}

\title{Colloidal shadows reveal hidden solute transport}
\author{Haoyu Liu}
\author{Zehao Chen}
\author{Amir A. Pahlavan}
\email{amir.pahlavan@yale.edu}
\affiliation{Mechanical Engineering and Materials Science, Yale University, New Haven, CT 06511, USA}

\begin{abstract}
An object releasing a chemical into a surrounding flow leaves behind a plume that encodes the rate of release, but for most solutes, that plume is invisible. Classical transport theory describes this configuration through a self-similar concentration boundary layer whose structure has never been resolved experimentally in forced flow at low Reynolds number: a century of measurements at these conditions has returned integrated transfer rates rather than the field itself. Here we show that both the field and the information it carries are experimentally accessible. Around hydrogel posts photopatterned inside a microfluidic channel, a released solute drives suspended colloidal particles away by diffusiophoresis, i.e., their drift along chemical gradients, carving a particle-free ``shadow'' downstream. Using a fluorescent solute, we resolve the concentration field around the post and recover the classical self-similar structure of convective mass transfer. For an invisible surfactant, the width of the shadow, benchmarked against a directly imaged solute on the same platform, then serves as a proxy for the dimensionless rate of mass transfer from the post. Finally, comparing two orientations of a triangular post tests a classical prediction by Brenner: the local concentration fields rearrange, yet no orientation dependence of the integrated transfer rate is resolved. For solute--tracer pairs whose diffusiophoretic response is known, colloidal shadows thus turn ordinary tracer particles into quantitative reporters of chemical exchange that cannot be observed directly.
\end{abstract}
\keywords{colloids, diffusiophoresis, microfluidics, low Reynolds number hydrodynamics, transport phenomena}
\maketitle

\section{Introduction}

How fast does a small object exchange dissolved chemicals with the fluid flowing past it? At low Reynolds number $Re \equiv (\textrm{inertial forces}/\textrm{viscous forces})$, and high P\'eclet number $Pe \equiv (\textrm{advective transport}/\textrm{diffusive transport})$, i.e., the regime of micrometric bodies in slow liquid flows, the answer is classical: a thin solute boundary layer wraps the body, forming a self-similar solute structure, and sets the dimensionless transfer rate, the Sherwood number $Sh \equiv (\textrm{total flux}/ \textrm{diffusive flux})$, through the scaling $Sh\propto Pe^{1/3}$~\citep{levich1962,friedlander1957,acrivos1962,frankel1968,batchelor1979,leal2007,yariv2024,subramanian2026}. That boundary layer sets two things at once: the integrated transfer rate, and how the transfer is distributed over the body's surface. The distribution is what a nearby organism actually experiences~\citep{stocker2012}, what carves a dissolving or melting solid into its terminal shape~\citep{huang2015,ristroph2012,huang2020,ristroph2018}, what shapes the release profile of a drug-loaded material~\citep{siepmann2012,limooney2016}, and what sets the chemical wake of a sinking organic aggregate~\citep{kiorboe2001,alcolombri2021}.

That distribution, however, is almost never observable. Slowly settling aggregates of organic matter in the ocean, i.e., marine snow, leak dissolved compounds while the microbial communities colonizing them respire, drawing oxygen down within and around the aggregate, and the plumes they trail shape microbial encounters and how much fixed carbon is remineralized near the surface rather than exported to depth~\citep{kiorboe2001,stocker2008,stocker2012,karthauser2021,raina2022}. Small aggregates with radii of a few hundred microns, settling at tens to hundreds of microns per second sit in the same low-$Re$, high-$Pe$ corner studied here~\citep{desai2019}, although larger particles settle faster and reach $Re\sim1$--$10$~\citep{moradi2018}. Oxygen is the exception that proves the rule, reachable through a dedicated sensor chemistry, including oxygen-sensitive tracer particles that image concentration and flow simultaneously around model aggregates~\citep{ploug1999,ahmerkamp2022}, but that route must be built analyte by analyte. Fluorescent labeling alters the chemistry it is meant to report, and interferometry resolves unlabeled fields only with refractive-index contrast and specialized optics~\citep{vogus2015,peshkovsky2026}. For most dissolved species the field that governs exchange is simply invisible.

\begin{figure*}[t]
\centering
\includegraphics[
    width=\textwidth,
    height=0.55\textheight,
    keepaspectratio,
    trim={15mm 0mm 15mm 0mm},
    clip,
    page=1
]{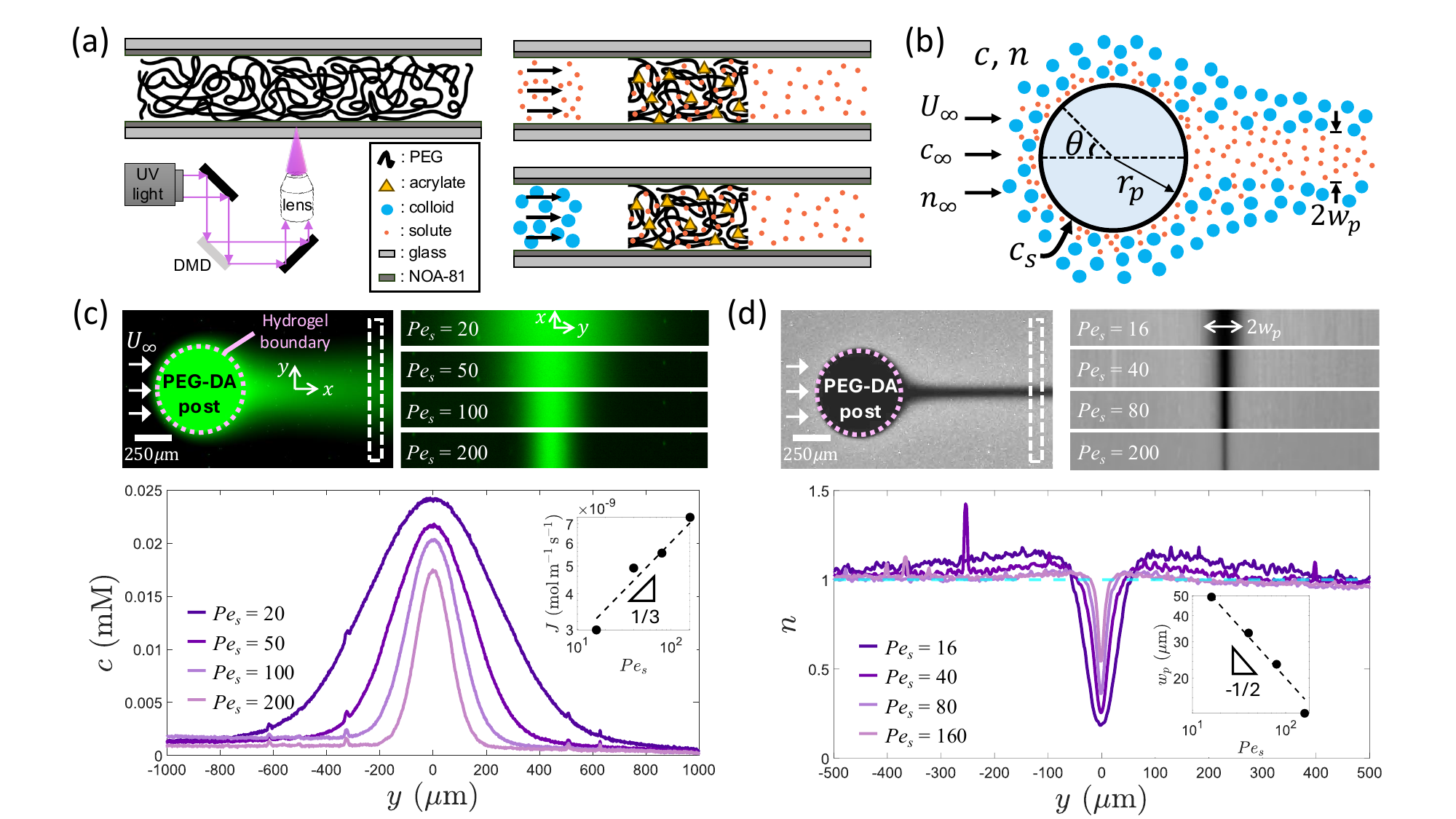}
\caption{\textbf{The solute plume emerging from a hydrogel post leaves behind a ``colloidal shadow''}. (a) Side view of the experimental setup. Left: a digital micromirror device (DMD) triggers hydrogel polymerization with designed patterns. Right: the hydrogel is loaded with solute (red dots) and self-releases it when the surrounding fluid has a lower concentration, while colloidal particles (blue dots) are introduced into the background flow. (b) Top view. Released solute induces diffusiophoretic drift of the colloids, forming a particle-free wake. The far-field solute and particle concentrations $c_\infty$ and $n_\infty$ as $r\to\infty$, surface solute concentration $c_s$ at $r=r_p$, and background flow velocity $U_\infty$ are indicated. (c) Fluorescence intensity field of fluorescein (upper left). The cross-tail intensity profile (upper right) yields the local concentration distribution (lower), whose integration gives the total solute flux per unit depth, exhibiting a $1/3$ power-law scaling with $Pe_s$ (lower inset). (d) Fluorescence intensity field of a carboxylate-modified particle suspension (200~nm diameter), with non-fluorescent SDS released from the hydrogel post (upper left). The depletion-tail width $2w_p$, defined by the two intersections of the particle concentration profile with $\tilde{n}=1$, follows a $-1/2$ power-law scaling with $Pe_s$ (lower inset).}
\label{fig1}
\end{figure*}

Colloidal particles offer a way to reveal it. Suspended particles drift along gradients of dissolved species, a phenomenon known as diffusiophoresis~\citep{deryagin1961,anderson1982,prieve1984,ebel1988,staffeld1989,anderson1989,abecassis2008,palacci2010,brady2011,palacci2012,khair2013,brady2021,velegol2016,shin2016,marbach2019,gupta2020,shim2022,ault2025}, and therefore rearrange wherever a solute is released. A releasing object thus casts what we call a \textit{colloidal shadow}, and reading transport from its geometry, \textit{colloidal shadowgraphy}, is the method we develop here. Prior work established colloidal rearrangement as a signature of solute release: soluto-inertial ``beacons'' attract or deplete particles over hundreds of microns in quiescent suspensions~\citep{banerjee2016,banerjee2019,banerjee2019design}; sedimenting beacons and dissolving objects leave comet-like trails, whose particle redistributions and integrated content have been measured and quantitatively modeled~\citep{banerjee2020,moller2023}; particle motion has revealed the collapse of sedimenting microgels~\citep{nowbahar2019};  engineered arrangements of solute sources and sinks have been used to sculpt colloidal density patterns in two dimensions~\citep{raj2023}; and hydrogel beacons have guided colloids to targets hidden inside porous media~\citep{tan2021}. These studies established predictable particle responses to imposed solute fields, but left the near-body field of a releasing body unresolved. Here, we present a fixed-body platform with externally controlled flow, calibrating the shadow width as a readout of the transfer rate.

This platform yields three results. With a fluorescent solute we resolve the near-body field and recover its self-similar structure and the $Pe_s^{1/3}$ scaling; with an invisible surfactant we calibrate the colloidal shadow as a proxy for the dimensionless transfer rate; and, comparing opposed orientations of a triangular post, we resolve no change in the integrated transfer, consistent with a classical invariance due to Brenner~\citep{brenner1967,brenner1970}. The two roles the particles play should be kept distinct: the spatial pattern responds to the hidden gradient field, while the shadow width at a fixed downstream station serves as a calibrated proxy for the dimensionless integrated transfer rate.

\begin{figure*}[t]
\centering
\includegraphics[
    width=\textwidth,
    trim={15mm 45mm 15mm 45mm},
    clip,
    page=2
]{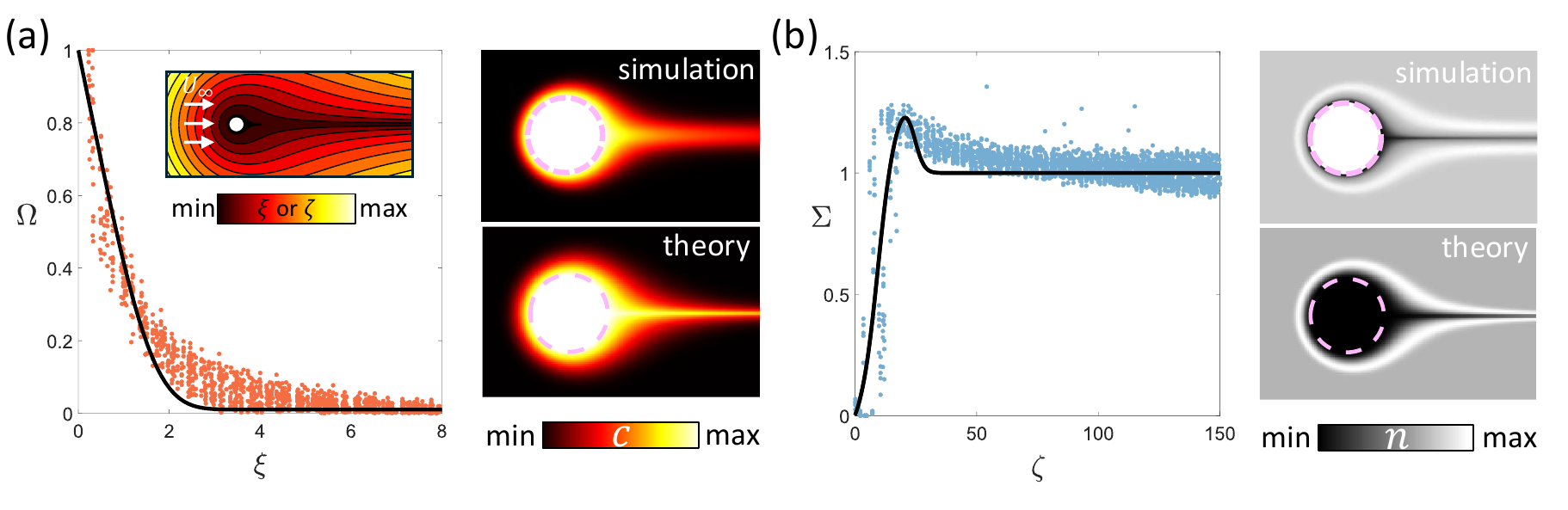}
\caption{\textbf{Self-similar structure of the solute and particle fields}. (a,b) Experimental concentration profiles (symbols) compared with the corresponding self-similar solutions (curves) for the solute (a) and the repelled diffusiophoretic particles (b), Eqs.~(\ref{soluteODE}) and~(\ref{particleODE}). Inset in (a): contour map of the self-similar coordinate $\xi$ (or $\zeta$). Right panels: numerical simulations alongside the theoretical solutions reconstructed in Cartesian coordinates.}
\label{fig2}
\end{figure*}

\section{A self-releasing hydrogel post in crossflow}

In a microfluidic channel, we fabricate a self-releasing cylindrical post by \textit{in-situ} UV polymerization of a poly(ethylene glycol) diacrylate (PEGDA) hydrogel~\citep{paustian2013,decock2018} using a digital micromirror device (DMD); an arbitrary cross-sectional shape can be programmed through the mirror array (Fig.~\ref{fig1}a and \textit{Materials and Methods}). The porous hydrogel absorbs solute from the surrounding solution, with a partitioning strength tunable by porosity and chemical affinity~\citep{banerjee2019design}. After equilibration, replacing the external fluid with a lower-concentration solution triggers gradual release, establishing a long-lasting gradient, the release mode pioneered with hydrogel membrane microwindows and soluto-inertial beacons~\citep{paustian2013,banerjee2016} and since used to hold steady electrolyte gradients in quiescent chambers~\citep{williams2024}. When the solute is fluorescent (here, fluorescein), the calibrated intensity distribution yields the external concentration field directly (Fig.~\ref{fig1}c). For solutes without such optical access, we add fluorescent diffusiophoretic tracers: here, the hydrogel is loaded with the non-fluorescent ionic surfactant sodium dodecyl sulfate (SDS), which repels negatively charged carboxylated polystyrene colloids, carving a particle-depletion tail of width $2w_p$ downstream of the post (Fig.~\ref{fig1}d). Although the release is intrinsically transient, the reservoir drains slowly (release timescale $\tau_p\sim10^2$--$10^3$~s) compared with the transit time to the measuring station ($x_{\rm obs}/U_\infty\lesssim30$~s), so each snapshot is quasi-steady (ratio $\lesssim6\times10^{-2}$ at the measuring station; \textit{SI Appendix}, \S3).

We non-dimensionalize as $(\tilde{r},\tilde{c},\tilde{n},\tilde{\bm{u}})=\left(r/r_p,\,\frac{c-c_\infty}{c_s-c_\infty},\,n/n_\infty,\,\bm{u}/U_{\infty}\right)$, with tildes denoting dimensionless quantities (Fig.~\ref{fig1}b). The steady dimensionless advection--diffusion equation for the solute is
\begin{equation}
    \frac{1}{Pe_s}\nabla^2\tilde{c} = \tilde{\bm{u}}\cdot\bm{\nabla}\tilde{c},
\label{soluteEqn}
\end{equation}
with $\tilde{c}=1$ at $\tilde{r}=1$ and $\tilde{c}\to0$ as $\tilde{r}\to\infty$. The colloids obey
\begin{equation}
    \frac{1}{Pe_p}\nabla^2\tilde{n} = \tilde{\bm{u}}\cdot\bm{\nabla}\tilde{n} + \underbrace{\frac{1}{Pe_{\textrm{dp}}}\bm{\nabla}\cdot\left[\tilde{n}\,\bm{\nabla}\ln(\tilde{c}+\gamma)\right]}_{\textrm{diffusiophoresis}},
\label{particleEqn}
\end{equation}
with a Robin (flux-balance) condition $D_p\,\partial\ln\tilde{n}/\partial\tilde{r}=\Gamma_p\,\partial\ln(\tilde{c}+\gamma)/\partial\tilde{r}$ at $\tilde{r}=1$ and $\tilde{n}\to1$ far away. Here $\gamma=c_\infty/(c_s-c_\infty)$ measures the background-to-release amplitude ratio ($\gamma\approx10^{-2}$ at early times in our experiments), and the P\'eclet numbers $Pe_s=U_\infty r_p/D_s$, $Pe_p=U_\infty r_p/D_p$, and $Pe_{\textrm{dp}}=U_\infty r_p/\Gamma_p$ compare advection with solute diffusion, particle diffusion, and diffusiophoretic drift (mobility $\Gamma_p$, negative for the repulsive pair used here~\citep{neryazevedo2017,ganguly2024}). Throughout, $Pe_s$ is evaluated with the diffusivity of the solute in question (\textit{SI Appendix}, Table~S1).

For the fluorescent solute, the release rate can be measured directly. All of the solute leaving the post is carried past any station downstream; in the far wake, where the velocity is nearly uniform and axial diffusion is negligible (\textit{SI Appendix}, \S2.3), integrating the measured concentration profile across the tail gives the total flux per unit depth,
\begin{equation}
    J = U_{\infty}\int_{-\infty}^{+\infty}\left(c-c_\infty\right)\,dy.
\label{wakeSurvey}
\end{equation}
Across flow speeds the data obey the classical high-$Pe$ scaling $J\propto Pe_s^{1/3}$ (Fig.~\ref{fig1}c, inset): a solute boundary layer of thickness $\delta_s\sim r_pPe_s^{-1/3}$ wraps the post, so $J\sim D_s(c_s-c_\infty)/\delta_s\propto Pe_s^{1/3}$. The invisible solute instead leaves its record in the particles: the depletion-tail half-width follows $w_p\sim r_pPe_s^{-1/2}$ (Fig.~\ref{fig1}d, inset), the signature of the diffusively spreading wake, where the flow is nearly parallel and the plume width grows as $\sqrt{D_sx/U_\infty}$~\citep{abecassis2008,shim2021,Temprano26}.

\section{Self-similar structure of the solute and particle fields}

At high $Pe_s$ the boundary layer admits a self-similar description, following classical analyses for the sphere~\citep{levich1962,leal2007}, spheroid~\citep{batchelor1979}, and cylinder~\citep{yariv2024}. In two dimensions, Stokes' paradox requires regularization of the near-field flow; following the hybrid asymptotic--numerical resolution~\citep{kaplunSaul1957,kropinski1995,yariv2024}, the near-surface streamfunction is $\psi = S(\chi)\left(\tilde{r}\ln\tilde{r}-\tilde{r}/2+1/(2\tilde{r})\right)\sin\theta$, with $\chi=(\ln Re^{-1}-1/2)^{-1}$ and $S(\chi)$ tabulated up to $Re\approx0.5$. Introducing the similarity variable $\xi=(\tilde{r}-1)Pe_s^{1/3}/f(\eta)$, with $\eta=\cos\theta$ and $f(\eta)$ the angular factor fixed by requiring similarity (\textit{SI Appendix}, \S2), Eq.~(\ref{soluteEqn}) collapses to
\begin{equation}
    \frac{d^2\Omega}{d\xi^2} + 3S(\chi)\,\xi^2\,\frac{d\Omega}{d\xi} = 0,
\label{soluteODE}
\end{equation}
where $\Omega(\xi)=\tilde{c}(\tilde{r},\theta)$, whose solution with $\Omega(0)=1$, $\Omega(\infty)=0$ is
\begin{equation}
    \Omega(\xi) = 1 - \frac{\int_{0}^{S^{1/3}\xi}\exp\left(-q^3\right)dq}{\int_{0}^{\infty}\exp\left(-q^3\right)dq}.
\label{soluteSoln}
\end{equation}
Replotted in $\xi$, the measured concentration fields collapse onto this master curve, and the solution reconstructed in Cartesian coordinates agrees with direct numerical simulation (Fig.~\ref{fig2}a; numerical methods in \textit{Materials and Methods}). The particle field inherits the same structure: in the stretched coordinate $\zeta=(\tilde{r}-1)Pe_p^{1/3}/f(\eta)$, Eq.~(\ref{particleEqn}) reduces to
\begin{equation}
    \Sigma'' = -3S(\chi)\zeta^2\Sigma' + \frac{Pe_p}{Pe_{\textrm{dp}}}\left[\left(\ln(\Omega+\gamma)\right)'\Sigma\right]',
\label{particleODE}
\end{equation}
with $\Sigma(\zeta)=\tilde{n}$, a Robin condition at the surface, and $\Sigma(\infty)=1$ (\textit{SI Appendix}, \S2). Its solution exhibits the experimentally observed structure, i.e., a depleted core, an accumulation peak where diffusiophoretic expulsion balances convection, and relaxation to the far field, onto which the data collapse (Fig.~\ref{fig2}b). The particle solution depends on the mobility ratio $\Gamma_p/D_p$, on $D_p/D_s$ through the $\zeta$--$\xi$ mapping, and on the amplitude ratio $\gamma$ (\textit{SI Appendix}, \S2), so the shape of the shadow itself carries information about the particle--solute pair.

\section{The colloidal shadow as a calibrated transport proxy}

\begin{figure}[t]
\centering
\includegraphics[
    width=\linewidth,
    trim={70mm 5mm 80mm 15mm},
    clip,
    page=3
]{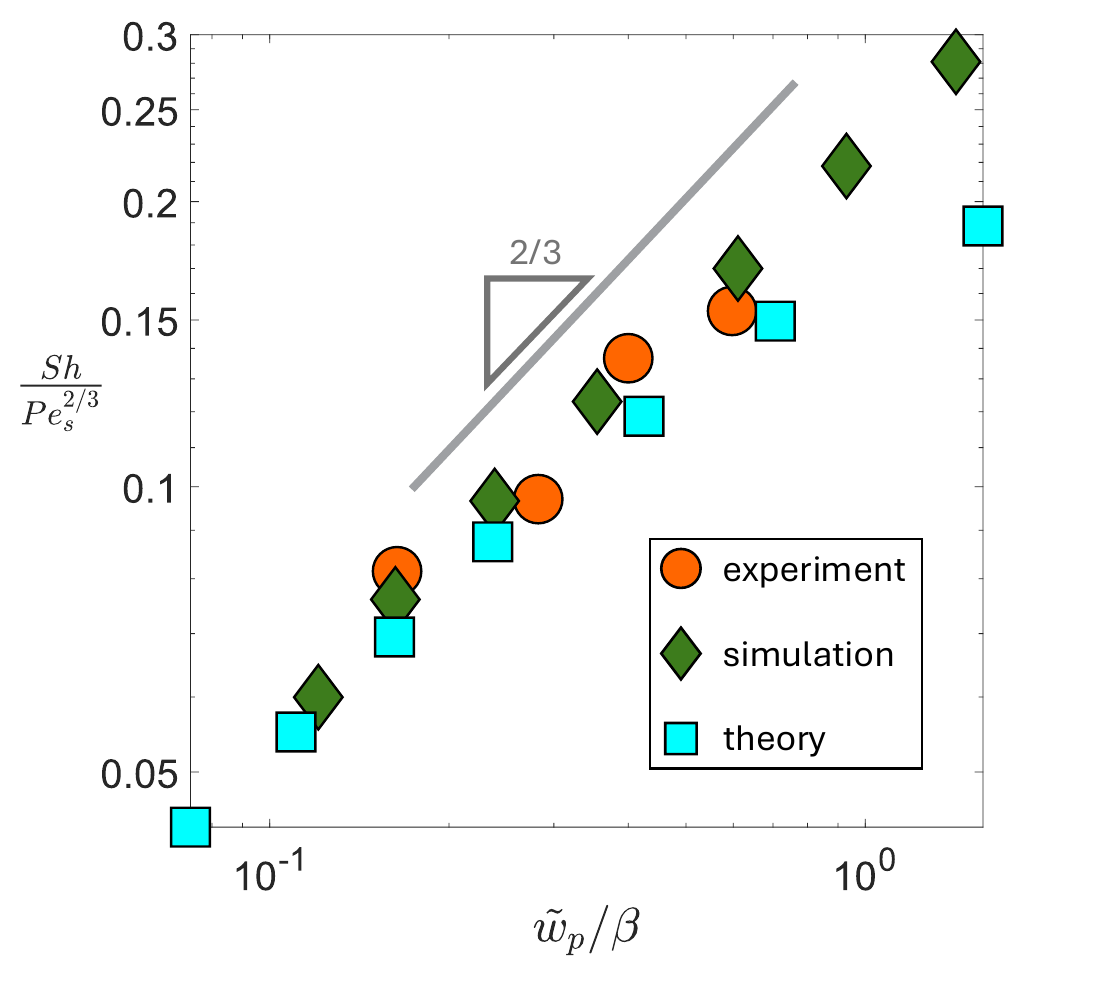}
\caption{\textbf{The colloidal shadow reports the dimensionless transfer rate of the hidden solute}. Rescaled transfer rate $Sh/Pe_s^{2/3}$ against the rescaled dimensionless shadow width $\tilde{w}_p/\beta$, where $\tilde{w}_p=w_p/r_p$: the calibrated transport map of Eq.~(\ref{fluxIndicator}). Experimental points combine $Sh$ measured from the fluorescein wakes with $\tilde w_p$ from SDS shadows; each quantity is nondimensionalized with its own solute's P\'eclet number (\textit{SI Appendix}, \S5). Numerical (diamonds) and theoretical (squares) results are shown alongside the gray line, which is a guide to the eye with slope $2/3$.}
\label{fig3}
\end{figure}

Having verified the forward physics, we turn to the inverse question: \textit{what does the shadow measure?} For a two-dimensional cylinder the dimensionless transfer rate is
\begin{equation}
    Sh = \frac{J}{2\pi D_s(c_s-c_\infty)},
\label{ShDef}
\end{equation}
and the similarity solution gives the transfer relation $Sh=\alpha Pe_s^{1/3}$, with $\alpha=0.580\,S^{1/3}$ for the cylinder, obtained by integrating the surface flux of the similarity solution, Eq.~(\ref{soluteSoln}), over the post surface~\citep{yariv2024}. The wake model (\textit{SI Appendix}, \S2.3) likewise gives the shadow width, $\tilde{w}_p\equiv w_p/r_p=\beta Pe_s^{-1/2}$, with $\beta$ set by the mobility ratio $\Gamma_p/D_s$, the amplitude ratio $\gamma$, and the measuring station. Linking the two observables we have
\begin{equation}
   \frac{Sh}{Pe_s^{2/3}} = \alpha\left(\frac{\tilde{w}_p}{\beta}\right)^{2/3}.
\label{fluxIndicator}
\end{equation}
Figure~\ref{fig3} shows experiment, simulation, and similarity theory falling together on the same curve, demonstrating that the shadow width serves as a \textit{calibrated proxy} for the dimensionless transfer rate. Two scalars are fitted here: the wake prefactor $\beta$, obtained from the width data, and the amplitude $c_s^{\rm (fluo)}$, set by the intercept (\textit{SI Appendix}, \S5). Everything else is independent: the two exponents, the shape of the similarity profile, and $\alpha$, known in closed form for the circle. The calibration proceeds as in hot-wire anemometry, where an electrical resistance is converted into a flow speed through an independently established law~\citep{king1914,Wood1968}: with the particle--solute mobility characterized independently~\citep{neryazevedo2017} and the wake prefactor $\beta$ fitted against the width data (\textit{SI Appendix}, \S\S2.3, 5), a measured width at known flow speed and post size maps through the transfer relation onto $Sh$.

\section{Flow reversal separates local structure from integrated transfer}

\begin{figure}[t]
\centering
\includegraphics[
    width=\linewidth,
    trim={60mm 0mm 60mm 0mm},
    clip,
    page=4
]{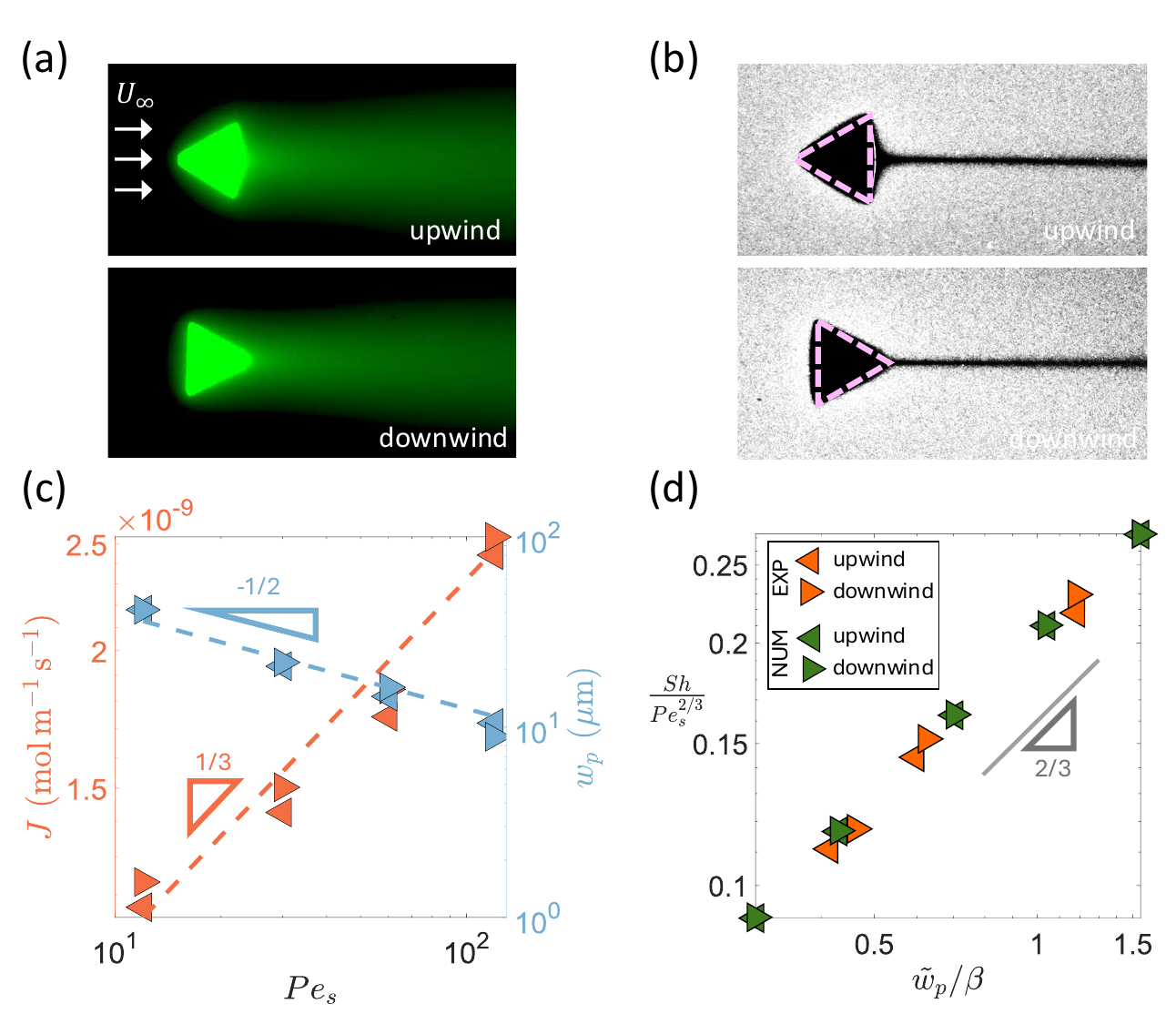}
\caption{\textbf{Flow reversal around a triangular hydrogel post leaves the integrated flux unchanged, as Brenner's classical flow-reversal theorem requires}. (a) Fluorescein intensity fields for the upwind (upper) and downwind (lower) orientations: the near-body fields rearrange with orientation. (b) Corresponding fields of the carboxylate-modified particle suspension (200~nm diameter), with non-fluorescent SDS released from the post. (c) Surface flux per unit depth $J$ and shadow width $w_p$ versus $Pe_s$: no difference between the two orientations is resolved across the full range. Blue and orange dotted lines are guides to the eye with slopes $-1/2$ and $1/3$ ($\triangleleft$: upwind; $\triangleright$: downwind). (d) $Sh/Pe_s^{2/3}$ against $\tilde{w}_p/\beta$ for both orientations; orange and green symbols denote experiments and simulations.}
\label{fig4}
\end{figure}

The local distribution of the flux is a quantity of consequence: it is what sculpts dissolving and melting bodies into their characteristic terminal forms~\citep{huang2015,ristroph2018,huangmoore2022}, and the carved shape is often the only record of the transfer the flow leaves behind~\citep{ristroph2018}. Both $J$ and $w_p$ are instead \textit{integral} measures: they say nothing about how the flux is distributed around the post. Flow reversal around an asymmetric body separates the local distribution from the integrated transfer. Reversing $U_\infty$ around a triangular post rearranges the local transfer, leading and trailing edges exchange roles and the near-body boundary layers redistribute, yet a theorem proved by Brenner for the same boundary-value problem, Eq.~(\ref{soluteEqn}), guarantees that the surface-integrated flux is invariant under $\bm{u}\to-\bm{u}$ in Stokes and potential flows, for a body of arbitrary shape~\citep{brenner1967,brenner1970}. This invariance is distinct from the kinematic reversibility of the flow itself, demonstrated with passive tracers in Hele--Shaw cells~\citep{flekkoy1996,oxaal1994}: it constrains an integral of a \textit{diffusing} field, requires only that the body surface be impermeable and of uniform concentration~\citep{vandadi2016,masoud2019}, and extends to transient release provided the surface history is the same in both flow directions~\citep{morrison1981}.

We test this symmetry with two triangular posts fabricated and loaded by nominally identical protocols, one oriented upwind and one downwind. The direct fluorescein measurement demonstrates that the near-body fields differ visibly between the orientations (Fig.~\ref{fig4}a), as they must for an asymmetric body, yet the measured fluxes $J$, obtained by integrating across the wake as in Eq.~(\ref{wakeSurvey}), agree between the two orientations across the full range of $Pe_s$ (Fig.~\ref{fig4}c). To our knowledge, these measurements constitute the first experimental comparison of integrated transfer between opposed flow orientations for mass transfer from a body in external flow, and they are consistent with Brenner's invariance (\textit{SI Appendix}, \S6). The reversal is equally a test of the colloidal probe itself. The particle fields rearrange visibly with orientation (Fig.~\ref{fig4}b), mirroring the solute's near-body reorganization; yet the shadow widths $w_p$ extracted from them show no resolvable difference between the two flow directions (Fig.~\ref{fig4}c) and, read through the calibrated map of Eq.~(\ref{fluxIndicator}), report the same transfer rate in both orientations (Fig.~\ref{fig4}d). Here a single wake prefactor $\beta_\triangle$, fitted to the combined triangular-post measurements, is used for both orientations.  The shadow therefore responds to the rearranging local field while reporting an integral that does not change with orientation.

\section{Discussion}

We have shown that a suspension of colloidal particles can serve as a calibrated reporter of hidden solute transport. On the benchmark of forced convection from a cylinder at low $Re$, the platform resolves the near-body concentration field and its high-$Pe$ self-similar structure, to our knowledge the first field-resolved observation of this structure in forced flow at $Re\ll1$, here in a confined microchannel realization, and a flow-reversal experiment on triangular posts finds the concentration fields rearranging while no orientation dependence of the integrated transfer is resolved, consistent with Brenner's invariance. The colloidal shadow of an invisible solute, analyzed through the same similarity theory, serves as a calibrated proxy for the solute's dimensionless transfer rate. The question we began with, i.e., how fast a small object exchanges chemicals with the flow past it, has had a theoretical answer for a century, and an experimental one only in the form of overall transfer coefficients. Those coefficients report how much is transferred; the colloidal shadow exposes how that transfer is organized, without labeling or directly imaging the solute, and does so even when the integral itself does not change.

Here, we have worked with repulsive particle--solute pairs, for which the shadow has a sharp, readily quantified edge; attractive pairs instead concentrate particles into bright accumulation features~\citep{banerjee2020}, whose inversion is less direct. The platform presumes a responsive particle--solute pair and a loadable gel: electrolytes and ionic surfactants drive strong diffusiophoresis, whereas non-electrolyte gradients drive weaker responses~\citep{anderson1982,williams2020,monier2025}, and reversible loading with a sufficient partition coefficient depends on the hydrogel's chemistry~\citep{paustian2013,banerjee2016,banerjee2019design}. Gel design is thus itself a decisive ingredient of the method, and an active one: solute-absorbing hydrogel boundaries have recently been shown to extend and prolong phoretic transport into dead-end pores~\citep{shah2026}. 

The thermal analogue of this problem was measured exhaustively by a century of hot-wire anemometry~\citep{king1914,hughes1916,hilpert1933,collis1959,Wood1968}, but in the form of overall heat-transfer coefficients, and in gases the P\'eclet number tracks the Reynolds number, so those experiments cannot reach $Pe\gg1$ while remaining at $Re\ll1$. In liquids, where $Pe\gg1$ is easily reached, slow flows are readily disturbed by density gradients: the mid-century wire measurements at $Re\lesssim1$ returned integrated rates~\citep{dobry1956,vogtlander1963}; dissolution measurements began only at $Re\sim10$--$500$ and included an additional contribution from natural convection that obscured the forced-convection limit~\citep{linton1960,steinberger1960}; sublimation and electrochemical techniques operate at moderate $Re$ or return surface-integrated rates~\citep{goldstein1995,selman1978}; and the one-third exponent itself has been confirmed only through overall transfer coefficients~\citep{harriott1962}. Even recent measurements on soluble particles settling at low $Re$ and high $Pe$ return integrated rates rather than fields~\citep{he2024}. Micrometric confinement removes the buoyancy effects that constrained those experiments.

Because the hydrogel sources are photopatterned \textit{in situ}, their geometry, arrangement, porosity, and chemistry are all design variables, and the background composition can be tuned to trade contrast for amplitude sensitivity; the same measurement therefore extends to complex shapes, to arrays of interacting sources, and to time-dependent release. Such extensions connect directly to the porous and confined geometries in which diffusiophoretic colloid transport is now being mapped~\citep{alipour2026,pujari2026,pahlavan2026}. In reactive settings the coupling runs both ways: colloids driven up a dissolving mineral's own solute gradient aggregate into a shell that slows the dissolution which produced it~\citep{roman2025}. The advance is thus less a new application of diffusiophoresis than a new use of it: suspended particles become passive mechanical transducers of scalar transport that is otherwise inaccessible. With both the source geometry and the forcing under experimental control, local field structure and global exchange can be interrogated separately, as the flow reversal shows directly.

\section{Materials and Methods}
\subsection{Microfluidic devices and hydrogel posts}
Channels (height $H=50~\mu$m, width $W=4$~mm) were fabricated in Norland Optical Adhesive 81 (NOA81), molded from PDMS masters following Ref.~\citep{bartolo2008} with modifications that suppress the oxygen-inhibition layer; NOA81's low oxygen permeability enables precise in-channel photopolymerization (\textit{SI Appendix}, \S1.1). A precursor of PEGDA ($M_n=400$) with 5\,\%\,v/v photoinitiator was injected and posts (circular radius $r_p=250~\mu$m; triangular cross-sections for the reversal experiments) were polymerized \textit{in situ} by DMD-patterned UV exposure on an inverted microscope (\textit{SI Appendix}, \S1.2). Posts were loaded by flowing 1~mM solute solution (sodium fluorescein or SDS) for 40~min; release was triggered by switching to a 0.01~mM background.

\subsection{Solutions, imaging, and analysis}
Fluorescein concentration fields were obtained from fluorescence images through a photobleaching-corrected calibration; SDS experiments used 1\,\%\,v/v carboxylate-modified polystyrene particles (200~nm diameter) in 0.01~mM SDS, with the diffusiophoretic mobility of the pair characterized in Ref.~\citep{neryazevedo2017}; those measurements lie at and above the upper end of our released concentration range, so we treat $\Gamma_p$ as an effective constant absorbed into the wake calibration rather than as a value independently verified at our background composition (\textit{SI Appendix}, \S\S1.3, 2.3). Wake fluxes were computed from cross-tail concentration profiles via Eq.~(\ref{wakeSurvey}); shadow widths were measured at the station $x=6r_p$ as the half-distance between the crossings of the particle-concentration profile with $\tilde{n}=1$ (Fig.~\ref{fig1}d). Per-solute P\'eclet numbers and all parameters are collected in \textit{SI Appendix}, Table~S1.

\subsection{Numerical simulations}
Steady flow and transport were solved by finite elements (COMSOL) in a two-dimensional exterior domain of size $80r_p\times80r_p$ with 2.3 million elements, refined at the post to resolve the boundary layer; the solute obeys Eq.~(\ref{soluteEqn}) and the particles Eq.~(\ref{particleEqn}) with the Robin flux-balance condition. A three-dimensional simulation of the confined channel shows mid-plane and depth-averaged fields indistinguishable from the two-dimensional model (\textit{SI Appendix}, \S4). Similarity solutions and the wake model are derived in \textit{SI Appendix}, \S2; quasi-steadiness of the release is quantified in \textit{SI Appendix}, \S3; the reciprocal-theorem proof and its experimental applicability are given in \textit{SI Appendix}, \S6.

\subsection{Data, Materials, and Software Availability}
All study data are included in the article and/or SI Appendix, or will be available in public repositories upon publication of the manuscript.

\section*{Author contributions}
A.A.P. designed research; H.L. and Z.C. performed experiments; H.L. performed simulations; H.L. and A.A.P. developed theory and analyzed data; H.L. and A.A.P.  wrote the paper with input from Z.C.

\section*{Competing interests}
The authors declare no competing interest.

\clearpage
\bibliography{hydrogel}

\end{document}